\documentclass[amsmath,amssymb,showpacs]{revtex4}
\usepackage{graphicx}
\usepackage{bm}

\begin{document}
\newcommand{\uu}[1]{\underline{#1}}
\newcommand{\pp}[1]{\phantom{#1}}
\newcommand{\be}{\begin{eqnarray}}
\newcommand{\ee}{\end{eqnarray}}
\newcommand{\ve}{\varepsilon}
\newcommand{\vs}{\varsigma}
\newcommand{\Tr}{{\,\rm Tr\,}}
\newcommand{\pol}{{\frac{1}{2}}}
\newcommand{\ba}{\begin{array}}
\newcommand{\ea}{\end{array}}
\newcommand{\bea}{\begin{eqnarray}}
\newcommand{\eea}{\end{eqnarray}}
\newcommand{\sinc}{{\,\rm sinc\,}}

\title{Cavity-QED tests of representations of canonical commutation relations employed in field quantization}
\author{Marek Czachor}
\author{Marcin Wilczewski}%
\affiliation{%
Katedra Fizyki Teoretycznej i Metod Matematycznych\\
Politechnika Gda\'nska, Narutowicza 11/12, 80--952 Gda\'nsk,
Poland}
\begin{abstract}
Various aspects of dissipative and nondissipative decoherence of Rabi oscillations are discussed in the context of field quantization in alternative representations of CCR. Theory is confronted with experiment, and a possibility of more conclusive tests is analyzed.
\end{abstract}

\maketitle

\section{Cavity QED in different representations of CCR}

Although the notion of entanglement between atomic and electromagnetic degrees of freedom plays a central role in quantum computing architecture based on cavity QED
\cite{Cirac,Kimble0,Raimond,Walther0,Chapman,Fisher,Mundt,LY1,LY2}, the very concept of entanglement leads to conceptual difficulties if quantum vacuum comes into play (cf. uniqueness of the vacuum versus violation of the Bell inequality \cite{Werner}, problems with teleportation of quantum fields \cite{Molotkov}, ambiguous entanglement with vacuum 
\cite{Enk,PC}). One of the problems is that the electromagnetic field can be quantized in different representations of canonical commutation relations (CCR). As shown in \cite{PC} the degree of entanglement is a representation-dependent property, and it is not clear which representations are really physical. The problem is a part of a wider and ongoing discussion on different quantization paradigms \cite{Q}. 

Now, can the available experimental data distinguish between different representations of CCR? The answer is less obvious than one might expect. In this Letter we will try to clarify the status of some data from cavity QED, and then discuss possibilities of more definitive tests.

We first analyze  at a representation independent level the simple problem of Rabi oscillation of a two-level atom in an ideal cavity (for technicalities we refer to \cite{WC}). In the second step we take into account two types of decoherence that should occur in realistic experiments. Following \cite{Tombesi}
we distinguish between dissipative and nondissipative decoherence and model dissipation employing the results of \cite{Chough}. Then we compare theoretical predictions based on irreducible representations with the experimental data of the Paris group \cite{Haroche}. Our conclusions are basically consistent with both  \cite{Tombesi} and \cite{Chough}: The observed decoherence appears to be entirely of a nondissipative type, but it is not clear why the effect of dissipation is invisible. Perhaps the fact that a photon is with probability 1 absorbed by the atom at times separated by the Rabi period leads to a sort of Zeno effect. This point requires further experimental and theoretical studies, and is beyond the scope of the present paper. 

Assuming that Rabi oscillations indeed do not reveal observable damping due to energy dissipation we ask to what extent the experiment can distinguish between reducible and irreducible representations of CCR. In physical terms the question can be translated as follows: How many oscillators do we need to model quantum fields? The standard answer is that we need one oscillator per mode. We show that in reducible representations the data only set certain limitations on the number of oscillators, and this number is independent of the number of modes. 

Finally, we suggest that one should repeat the measurements reported in \cite{Haroche} with better cavities, finer time resolution, and monitor the Rabi oscillation for longer times. The point is that the decay due to experimental imprecisions may mask quantum beats of a completely new type and origin. In principle, the beats can be observed in a form of vacuum collapses and revivals, the effect occurring in reducible $N$-representations \cite{nCCR}. Observation of the revival would be of fundamental importance for our understanding of field quantization.

\section{Rabi oscillations at a representation-independent level}

Similarly to \cite{Haroche} we work with the Jaynes-Cummings model \cite{JC,Allen}. The crucial point is that we begin with solving Heisenberg equations of motion for the two-level atom at a representation independent level \cite{AR}. The CCR algebra, in its general form, reads
\be
[a_k,a^*_{k'}]
=\delta_{kk'} I_k,
\label{ccrrel}
\ee
$I_k$ commute with all the other operators, and $I_k^*=I_k$. We do not assume that $I_k$ is proportional to the identity (this generality will pay, as we shall see shortly). By Schur's lemma $I_k$ is necessarily proportional to the identity only in irreducible representations. 
We employ the usual notation \cite{Allen} where $R_l=\sigma_l/2$, $R_\pm=R_1\pm i R_2$, 
$\sigma_l$ are the Pauli matrices, and $g$ is a complex coupling parameter.
We assume there exists a free-field Hamiltonian $H_0$ satisfying
$[a_k,H_0]= \omega_k a_k$,
$[a_k^*,H_0]= -\omega_k a_k^*$.
Note that $H_0$ cannot, in general, be given by $\sum_k\omega_k a_k^*a_k$; the latter works only for some representations (e.g. for irreducible representations with $I_k$ equal to an identity, or for the reducible `$N=1$' representation; `$N>1$' reducible representations require a different construction). 

Let us now select a frequency $\omega_{p}=|\bm p|=\omega$ and assume that only this frequency couples to the two-level system. We also split $H_0$ into two parts: $H_0^\perp$ commuting with $a_p$ and $a_p^*$, and 
$H_0^\parallel=\omega N_p$, where
$[a_p,N_p]=  a_p$, $[a_p^*,N_p]= - a_p^*$. 
The model is given by the full Hamiltonian
\be
H
=
\omega_0 R_3
+
H_0
+
g R_+ a_p
+
\bar g R_- a^*_p.\label{H}
\ee
Solving the Heisenberg picture equations we find
\begin{widetext}
\be
R_3(t)
=
R_3
\bigg(
1
-
2|g|^2  X
\frac{\sin^2({\Omega}_R t)}{{\Omega}^2_R}
\bigg)
+
\bigg(
\frac{\Delta}{2}
\frac{\sin^2({\Omega}_R t)}{{\Omega}^2_R}
-
i
\frac{\sin(2{\Omega}_R t)}{2{\Omega}_R}
\bigg)
g R_+ a_p
+
\bigg(
\frac{\Delta}{2}
\frac{\sin^2({\Omega}_R t)}{{\Omega}^2_R}
+
i
\frac{\sin(2{\Omega}_R t)}{2{\Omega}_R}
\bigg)
\bar g R_- a^*_p,\nonumber\\
&{}&
\label{R3tgen}
\ee
\end{widetext}
where $\Delta=\omega_0-\omega$, $\Omega_R=\sqrt{\Delta^2/4+|g|^2 X}$, and
\be
X
&=&
(R_3+1/2) I_p
+
a_p^* a_p.\label{X}
\ee
The next important notion that can be introduced at a general level is the displacement operator
\be
D(z)
&=&
\exp\sum_k\big(z_k a^*_k-\bar z_k a_k\big).\label{D}
\ee
Acting with $D(z)$ on a vacuum vector we obtain a coherent state. Its form depends on what is meant by vacuum in a given representation. 

We will not discuss in more detail the irreducible representations since, as shown in  
\cite{WC}, they all yield physically equivalent and well known results. Instead, we directly turn to the `$N<\infty$' reducible representation introduced in \cite{1} and worked out in many details in \cite{2,3,4}. 

\section{$N<\infty$ representation}

The representation is constructed as follows. For simplicity we ignore here the polarization degree of freedom (see however \cite{1,2,3,4}). Take an operator $a$ satisfying $[a,a^*]=1$ and the kets $|\bm k\rangle$ corresponding to standing waves in some cavity. We define 
\be
a_k = |\bm k\rangle\langle \bm k|\otimes a,\quad
I_k = |\bm k\rangle\langle \bm k|\otimes 1.\label{I N=1}
\ee
The operators (\ref{I N=1}) satisfy (\ref{ccrrel}), where
$\delta_{kk'}$ is the 3D Kronecker delta. The fact that $I_k$ is not proportional to the identity means that the representation is reducible. In our terminology this is the `$N=1$ representation'. Its Hilbert space $\cal H$ is spanned by the kets 
$|\bm k,n\rangle=|\bm k\rangle|n\rangle$, where $a^*a|n\rangle=n|n\rangle$. Such a Hilbert space represents essentially a single harmonic oscillator of indefinite frequency (for physical motivation cf. \cite{1,2} and the Appendix in \cite{WC}). An important property of the representation is that  $\sum_k I_k=I$ is the identity operator in $\cal H$. 
A vacuum of this representation is given by any state annihilated by all $a_k$. The vacuum state is not unique and belongs to the subspace spanned by $|\bm k,0\rangle$. In our notation a $N=1$ vacuum state reads $|O\rangle=\sum_k O_k|\bm k,0\rangle$ and is normalized by $\sum_k |O_k|^2=\sum_k Z_k=1$, $Z_k=|O_k|^2$. Such a vacuum represents a single-oscillator ground-state {\it wavepacket\/}. As shown in 
\cite{2,3} in a fully relativistic formulation the maximal probability $Z=\max_k\{Z_k\}$ is a
Poincar\'e invariant and plays a role of renormalization constant. 
For $N\geq 1$ the representation space is given by the tensor power 
$\uu{\cal H}={\cal H}^{\otimes N}$, i.e. we take the Hilbert space of $N$ (bosonic) harmonic oscillators. Let $A: {\cal H}\to {\cal H}$ be any operator for $N=1$. 
We denote
$A^{(n)}=I^{\otimes (n-1)}\otimes A \otimes I^{\otimes (N-n)}$, 
$A^{(n)}: \uu{\cal H}\to \uu{\cal H}$, for $1\leq n\leq N$. For arbitrary $N$ the representation is defined by
\be
\uu a_k
&=&
\frac{1}{\sqrt{N}}\sum_{n=1}^N a_k^{(n)},\quad
\uu I_k=\frac{1}{N}\sum_{n=1}^N I_k^{(n)}, \label{IN}\\
{[\uu a_k,\uu a_{k'}^*]} &=& 
\delta_{kk'}\uu I_k,\quad
\sum_k \uu I_k=\uu I=I^{\otimes N}
\ee
and the $N$-oscillator vacuum is the $N$-fold tensor power of the $N=1$ case, a kind of Bose-Einstein condensate consisting of $N$ wavepackets:
\be
|\uu O\rangle=|O\rangle\otimes\dots\otimes|O\rangle=|O\rangle^{\otimes N}.\label{vacN}
\ee
The free-field Hamiltonian is, for $N=1$ and $\omega_k=|\bm k|$, 
\be
H_0=\sum_k \omega_ka_k^*a_k
=
\sum_k \omega_k|\bm k\rangle\langle \bm k|\otimes a^*a.
\ee
In each eigensubspace with fixed $|\bm k\rangle$ the operator $H_0$ is just an ordinary Hamiltonian of the oscillator with frequency $\omega_k$. 
Let us note that one can work also with 
\be
H_0=\frac{1}{2}\sum_k \omega_k(a_k^*a_k+a_k a_k^*)
=
\sum_k \omega_k|\bm k\rangle\langle \bm k|\otimes a^*a +\frac{1}{2}\sum_k \omega_k I_k.
\ee
The vacuum term is a well-defined Hermitian operator, and can be removed by a well-defined unitary transformation. This is an example of a procedure that can be termed, after Finkelstein, ``regularization by quantization" \cite{Q}.
For arbitrary $N$ the generator of free field evolution is the Hamiltonian of $N$ noninteracting oscillators, i.e. 
$\uu H_0=\sum_{n=1}^N H_0^{(n)}$.
Let us stress that $\uu H_0$ should not be confused with 
$\sum_k \omega_k\uu a_k^*\uu a_k$. The operator $\uu a_k^*\uu a_k$ nevertheless occurs in (\ref{X}) and thus plays an important role in the Jaynes-Cummings problem. Our definition of $\uu H_0$ implies that $[\uu a_k,\uu H_0]=\omega_k\uu a_k$ which is the formula we required at the representation independent level. 

A monochromatic coherent state with frequency $\omega$ is given by the usual formula
\be
|\uu z\rangle
&=&
\exp\big(z \uu a_p^*-\bar z \uu a_p\big)|\uu O\rangle\label{|beta>}.
\ee
Starting with the excited state and a vacuum field, 
$|{\Psi}\rangle=|+\rangle|\underline{O}\rangle$, we find 
\begin{widetext}
\be
w(t)
=
\langle {\Psi}|R_3(t) |{\Psi}\rangle
=
\frac{1}{2}
-
\sum_{s=0}^N
|g|^2\frac{s}{N}
\frac{\sin^2\sqrt{\Delta^2/4+|g|^2s/N }t}{\Delta^2/4+|g|^2s/N}
\left(
\begin{array}{c}
N\\s
\end{array}
\right)
Z_p^s (1-Z_p)^{N-s}.
\label{R3t_1}
\ee
\end{widetext}
So this is the vacuum Rabi oscillation in the reducible representation, and the last term is the binomial distribution for $N$ trials, with single-trial probability of success $Z_p$. There are $N$ different frequencies and thus collapses and revivals will necessarily occur if $1<N<\infty$. For $N$ large enough the binomial distribution can be approximated by a Gaussian, and one can show that for small $Z$ the  parameter that controls the Rabi oscillation is effectively the product $NZ$ (see below, Sec.~VIII). 
The limit $N\to\infty$ (with fixed $Z$) can be computed on the basis of the law of large numbers for the binomial distribution, 
\be
\lim_{N\to\infty}w(t)
=
\frac{1}{2}
-
|g|^2Z_p
\frac{\sin^2\sqrt{\Delta^2/4+|g|^2 Z_p }t}{\Delta^2/4+|g|^2 Z_p},
\label{R3t_1->}
\ee
i.e. the frequency $s/N$ approaches the probability of success in a single trial of the Bernoulli process, 
$s/N\to Z_p$. (\ref{R3t_1->}) is essentially the standard Jaynes-Cummings prediction, but with a modified coupling. It is clear that the measurable coupling is not just $g$ but rather its renormalized version $g_{\rm ph}=g\sqrt{Z}$.
Let us note that this is equivalent to bare charge renormalization: $e_{\rm ph}=e_0 \sqrt{Z}$. $Z$ is therefore an analogue of the renormalization constant $Z_3$ and $\chi_k=Z_k/Z$ plays a role of a cut-off \cite{cut-off}. 

Both the cut-off and the renormalization constant occur here automatically. If we assume that for optical frequencies 
$Z_p=\max_k\{Z_k\}=Z$ (i.e. $\chi_p=1$) the agreement between the irreducible case and the $N\to\infty$ limit of the reducible one is exact. 
The law of large numbers plays here a role of a correspondence principle with the standard formalism, a property not limited only to the Jaynes-Cummings example. 

With this background in mind one can easily generalize the discussion to thermal and coherent states, and mixed atomic initial condition \cite{WC}. Let $p_+$ and $p_-$ denote initial probabilities of finding the atom in, respectively, excited and ground states. Replacing vacuum by a thermal light with the distrubution
\be
{\tt P}(n)
&=&
\frac{\bar n^n}{(1+\bar n)^{(n+1)}},
\ee
we find, $p_+(t)=w(t)+1/2$,
\begin{widetext}
\be
p_+(t)
&=&
p_+
-
\sum_{s=0}^N
\left(
\begin{array}{c}
N\\s
\end{array}
\right)
Z_p^s (1-Z_p)^{N-s}
\sum_{n=0}^\infty
{\tt P}(n)
\Big(p_+-\frac{p_-\bar n}{1+\bar n}\Big)
|g|^2\frac{(n+1)s}{N}
\frac{\sin^2\big(t\sqrt{\frac{\Delta^2}{4}+|g|^2 \frac{(n+1)s}{N}} \big)}
{\Delta^2/4+|g|^2 (n+1)s/N}.
\ee
The limit $N\to\infty$
\be
p_+(t)
&=&
p_+
-
\sum_{n=0}^\infty
{\tt P}(n)
\Big(p_+-\frac{p_-\bar n}{1+\bar n}\Big)
|g_{\rm ph}|^2\chi_p
\frac{\sin^2\big(t\sqrt{\Delta^2/4+|g_{\rm ph}|^2 (n+1)\chi_p} \big)}
{\Delta^2/4+|g_{\rm ph}|^2(n+1)\chi_p},
\ee
is, up to $\chi_p$, known from irreducible representations.
For a coherent state $|\uu z\rangle$ and $p_+=1$ we find
\be
p_+(t)
&=&
1
-
\sum_{s=0}^N
\sum_{n=0}^\infty
\frac{|g_{\rm ph}|^2 (n+1)}{Z}\frac{s}{N}
\frac{\sin^2\big(t\sqrt{\Delta^2/4+|g_{\rm ph}|^2 (n+1)s/(ZN)} \big)}
{\Delta^2/4+|g_{\rm ph}|^2 (n+1)s/(ZN)}
\frac{|z\sqrt{\frac{s}{N}}|^{2n}}{n!}
e^{-|z\sqrt{\frac{s}{N}}|^2}
\left(
\begin{array}{c}
N\\s
\end{array}
\right)
Z_p^s (1-Z_p)^{N-s}\nonumber.
\ee
The limiting form, for $N\to\infty$, is again familiar
\be
\lim_{N\to\infty}p_+(t)
&=&
1
-
\sum_{n=0}^\infty
|g_{\rm ph}|^2 (n+1)\chi_p
\frac{\sin^2\big(t\sqrt{\Delta^2/4+|g_{\rm ph}|^2 (n+1)\chi_p} \big)}
{\Delta^2/4+|g_{\rm ph}|^2 (n+1)\chi_p}
\frac{|z_{\rm ph}\chi'_p|^{2n}}{n!}
e^{-|z_{\rm ph}\chi'_p|^2}.
\ee
\end{widetext}
For the same reason as before we obtain the standard formula but with the cut-offs $\chi_p=Z_p/Z$, 
$\chi'_p=\sqrt{\chi_p}$, and renormalized $e_{\rm ph}=e_0\sqrt{Z}$, 
$z_{\rm ph}=z\sqrt{Z}$.  

\section{Dissipative and nondissipative decoherence}

An analysis of realistic experiments must take into account decoherence. There are two main sources of decoherence that were identified in the literature in the context of the experiment of Brune {\it et al.\/} \cite{Haroche}. 

The analysis of dissipation based on quantum trajectories approach \cite{Chough,Chough2} leads to the conclusion that the damping due to energy loss in the cavity should have the form 
$p_{\kappa,+}(t)=e^{-\kappa t}p_+(t)$, where $p_+(t)$ is the probability of finding the atom in the excited state in an ideal cavity, and $2\kappa=1/T_{\rm cav}$. The factor 2 takes into account the fact that energy is not dissipated if the atom is in the excited state and there is no photon in the cavity. Obviously, for $t\to\infty$ the atom is with certainty found in its ground state. 

The second source of decoherence is nondissipative in nature and was discussed in 
\cite{Tombesi}. It originates from the fact that the data collected at time $t$ should not be compared directly with $\rho(t)$ describing   the state computed on the basis of first principles, but with the average
\be
\rho_{\Delta t}(t)
&=&
\int_0^\infty dt'\,p_{\Delta t}(t,t')\rho(t'),
\ee
where $\rho(t')$ is the first-principles state and $p_{\Delta t}(t,t')$ describes our lack of knowledge as to the exact duration of time evolution. The data from \cite{Haroche} involve a sample of 90 points selected from the time interval $0<t<90\,\mu$s. Therefore the time-of-measurement uncertainty may be assumed to satisfy $0<\Delta t<1\,\mu$s, which indeed turns out to reasonable model the data. However, it is not evident if this is really the true explanation of the discrepancy. The problem is that another value of $\Delta t$ is also mentioned  in 
\cite{Tombesi}, namely $\Delta t=0.01 t$. It would lead to a linear growth of $\Delta t$ between $0.01\,\mu$s and $0.9\,\mu$s, and then the agreement between theory and experiment is worse. 

Nevertheless, leaving aside this and similar subtleties, we may use the probability distribution introduced in \cite{Tombesi}
\be
p_{\Delta t}(t,t')
&=&
\frac{e^{-t'/\Delta t}}{\Delta t}\frac{(t'/\Delta t)^{t/\Delta t-1}}{\Gamma(t/\Delta t)}.
\ee
In all the representations discussed in this paper we have arrived at atomic probabilities involving terms of the form
\be
p_+(t)
&=&
A+B\sin^2 \Omega t.
\ee
The associated effective probabilities then read
\begin{widetext}
\be
p_{\Delta t,\kappa,+}(t)
&=&
\int_0^\infty dt'\, p_{\Delta t}(t,t')
e^{-\kappa t'}(A+B\sin^2 \Omega t')
\nonumber\\
&=&
(1+\kappa\Delta t)^{-t/\Delta t}
\Bigg[
A+\frac{1}{2}B\Bigg(1
-
\Big[1+\Big(\frac{2\Omega\Delta t}{1+\kappa\Delta t}\Big)^2\Big]^{-\frac{t}{2\Delta t}}
\cos\Big(\frac{t}{\Delta t} \arctan\frac{2\Omega\Delta t}{1+\kappa\Delta t}\Big)
\Bigg)
\Bigg].\label{22}
\ee
\end{widetext}
The overall damping factor $(1+\kappa\Delta t)^{-t/\Delta t}$ is the deformed exponential \cite{Naudts} occurring in non-extensive thermodynamics \cite{Tsallis}, and whose links to Gamma-function averages are well known \cite{Wilk}.

\section{Experiment of the Paris group --- vacuum Rabi oscillation} 

Let us first concentrate on the vacuum Rabi oscillation observed in \cite{Haroche}. 
This part of the data is particularly intriguing and plays an important role for calibration of the experimental setup. Theoretical fits shown in \cite{Haroche} were based on sinusoids exponentially damped by $\exp (-t/T)$ with $T=40\,\mu$s \cite{Raimond1}. It is essential that the parameter $T$ was much smaller from the reported value $T_{\rm cav}=220\,\mu$s of the cavity lifetime. The coupling constant employed in the fits was $g_{\rm ph}/\pi=47$ kHz, and the cavity was filled with $0.8$~K thermal light (average number of photons $\bar n=0.05$). Brune {\it et al.\/} tried to explain the difference between $T$ and $T_{\rm cav}$ by means of dark counts and collisions with background gas.

The role of dark counts and collisions was analyzed in detail in \cite{Chough}, but the conclusion was negative --- the source of the discrepancy had to be different. The analysis presented in 
\cite{Chough} revealed also another problem with the data: The cavity lifetime $T_{\rm cav}=220\,\mu$s should induce a shift of excited-state probability towards zero (as in Fig.~1, left part), but there is no trace of this phenomenon. 

Let us now turn to the solution of the discrepancy between  $T$ and $T_{\rm cav}$ in terms of nondissipative decoherence, suggested in \cite{Tombesi}. In Fig.~1 we show the prediction involving both kinds of decoherence, and based on irreducible representations of CCR. The initial probability is $p_+=0.99$. For the left plots the damping factor is 
$\kappa=1/(2 T_{\rm cav})=10^6/440$~Hz, and three different values of $\Delta t$ are compared. The value $\Delta t \approx 0.5\,\mu$s (suggested in \cite{Tombesi}) would reasonably reproduce the data if one neglected the energy decay, a fact consistent with the observations from \cite{Chough}. In the right plots $\kappa=0$. For times $0<t<15 \,\mu$s the data are then consistent with $\Delta t=0.7\,\mu$s. The next peak is well described by $\Delta t=0.3\,\mu$s, then again the minimum looks like $\Delta t=0.7\,\mu$s, and finally we can use $\Delta t=0.5\,\mu$s. 
\begin{figure}
\includegraphics[width=16 cm]{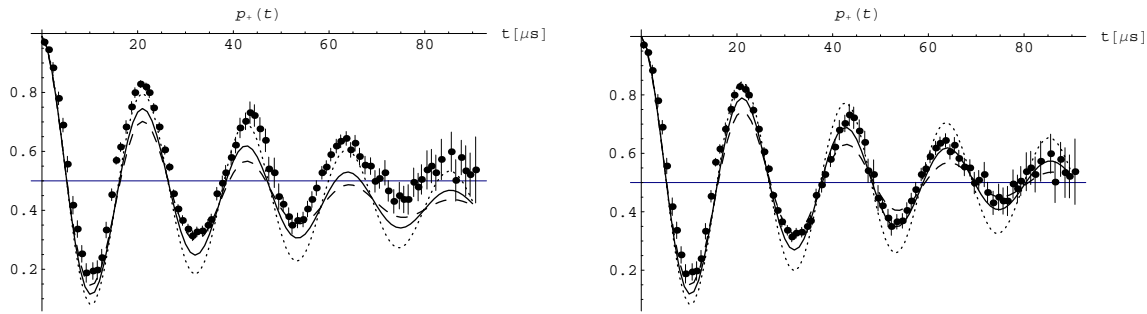}
    \caption{Standard theory of $p_+(t)$ for three different uncertainties of atomic time-of-flight measurements: $\Delta t=0.3\,\mu$s (dotted), $\Delta t=0.5\,\mu$s (full, suggested in \cite{Tombesi}), $\Delta t=0.7\,\mu$s (dashed). Left plots: $T_{\rm cav}=220\,\mu$s, $\kappa=1/(2T_{\rm cav})$. Right plots: $\kappa=0$.}
\end{figure}
The fits are quite sensitive to small variations of $\Delta t$, but generally the solution is acceptable if one could explain why $\kappa=0$ is here meaningful.
The problem is a serious one since the whole logic of the experiment is based on nonnegligible dissipation (cavity decay eliminates the maser efect). 

\section{Experiment vs. $N<\infty$ representations}

Let us now turn to the case of $N<\infty$ representations. The first question we have to clarify is what is the role of $\kappa$ for predictions based on reducible representations. In Fig.~2 (left) we compare the standard prediction from Fig.~1 for $T_{\rm cav}=220\,\mu$s and $\Delta t=0.5\,\mu$s (dotted) with an analogous result for the reducible $N=2000$, $Z=0.1$ representation (full). The two curves differ by less than experimental error bars, and it is clear that the finite-$N$ representations suffer from the same problem as the irreducible ones: For $\kappa=1/(2T_{\rm cav})>0$ the Lindblad-type plots are shifted downwards with respect to the data. 
This is not surprising, since for $N\to\infty$ the reducible representation should reconstruct predictions of the irreducible one. 

In Fig.~2 (right) we show the same situation as in Fig.~2 (left) but now with $\kappa=0$. The data are consistent with $NZ=200$. It is very important to keep in mind that for $NZ>200$ the agreement between the two theories will be even better. This is why this type of experiment will not be able produce an exact value of $NZ$, but only set a lower bound on the value of this parameter. Fig.~3 shows analogous plots for coherent states. 
Collecting all the avaliable data we can estimate a common lower bound following from various experimental situations --- here in all the plots the lower bound $NZ>200$ is enough to have predictions experimentally indistinguishable from the standard theory. 
Possibility of a test directly determining $NZ$ is discussed in the next section.

\begin{figure}
\includegraphics[width=16 cm]{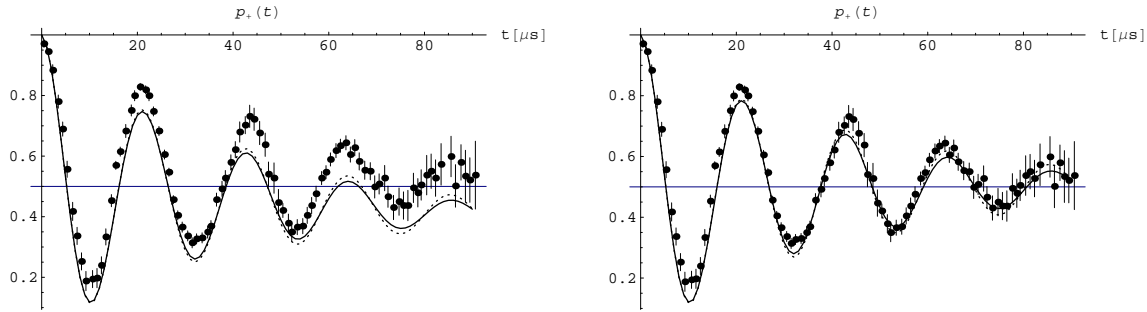}
    \caption{Comparison of the standard theory of $p_+(t)$ (dotted) with the reducible representation characterized by $N=2000$, $Z=0.1$ (full). The left plots, 
    $\kappa=1/(2T_{\rm cav})=10^6/440$~Hz, are shifted downwards with respect to the data. The right plots employ $\kappa=0$. All the curves correspond to $\Delta t=0.5\,\mu$s. Predictions of the two theories differ by less than experimental error bars.}
\end{figure}

\section{Can we directly measure $NZ$?}

Reducible and irreducible representations are idistinguishable as long as the beats typical of finite $N$ are masked by the decay caused by a nonzero $\Delta t$. In Fig.~4 we show the dynamics of $p_+(t)$ monitored with the resolution $\Delta t=0.005\,\mu$s. 
We assume that initially the atom is in the upper level and there are no photons (exact vacuum state at zero temperature). The plots reveal two important features of finite-$N$ representations. First of all, even in exact vacuum we find beats analogous to what is known from irreducible-representation coherent states. Secondly, the first revival occurs after a time that depends effectively on the product $NZ$, and not separately on $N$ and $Z$. To understand why this has to happen we replace the binomial distribution by its asymptotic  form, valid for large $N$,
\be
\left(
\begin{array}{c}
N\\s
\end{array}
\right)
Z_p^s (1-Z_p)^{N-s}
\approx
\left(
\begin{array}{c}
N\\s
\end{array}
\right)
Z^s (1-Z)^{N-s}
\approx
\frac{e^{-\frac{(s-NZ)^2}{2NZ(1-Z)}}}{\sqrt{2\pi NZ(1-Z)}}
\approx
\frac{e^{-\frac{(s-NZ)^2}{2NZ}}}{\sqrt{2\pi NZ}}.
\ee
The shape of the Gaussian is controlled mainly by the product $NZ$. The smaller $Z$ the less important its exact value (the last approximate equality holds for small $Z$). Increasing 
$N$ with $Z$ kept constant we shift the first revival more to the right. In the limit $N\to\infty$ the first revival is shifted to infinity, and we recover the standard undamped oscillation. So, the absence of the revival in an experiment can only set a lower bound on 
$NZ$, and is not a proof that the physical representation is irreducible.

\begin{figure}
\includegraphics[width=16 cm]{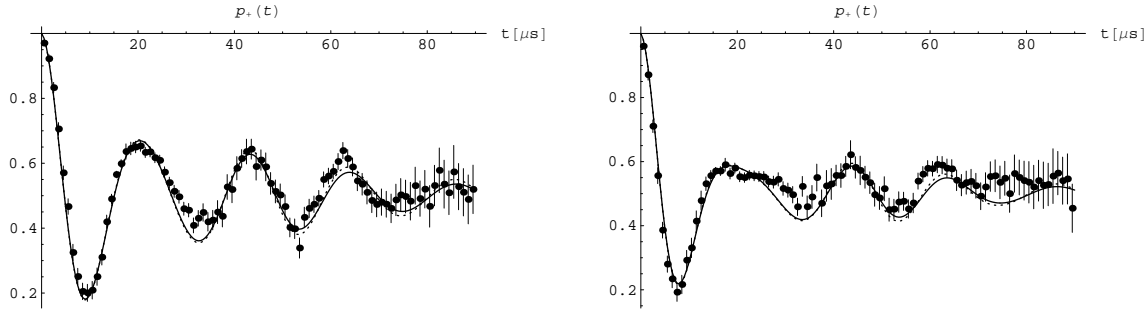}
    \caption{Comparison of the standard theory of $p_+(t)$ (dotted) with the reducible representation characterized by $N=2000$, $Z=0.1$ (full). Coherent states with $\bar n=0.4$ (left),  $\bar n=0.85$ (right), and $\kappa=0$, $\Delta t=0.5\,\mu$s. Initially the atom is in the upper state $(p_+=1)$.}
\end{figure}
\begin{figure}
\includegraphics[width=16 cm]{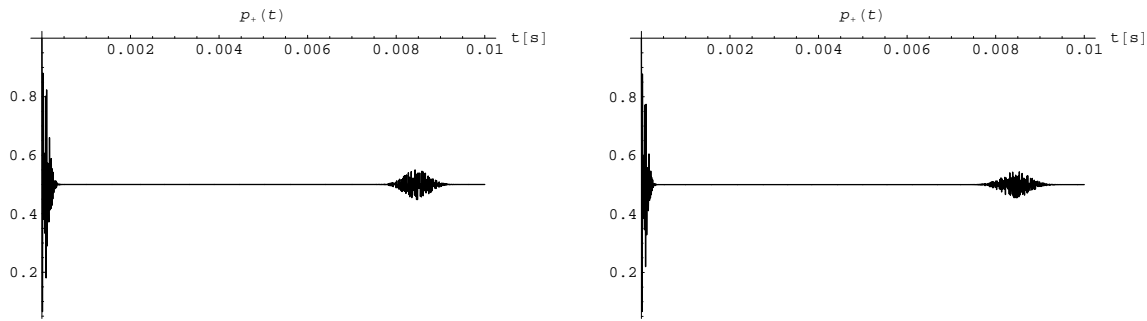}
    \caption{Reducible representations for $NZ=200$, $\kappa=0$, and $\Delta t=0.005\,\mu$s. Vacuum Rabi oscillation is monitored for a longer time. Left plot: $N=600$, $Z=1/3$. Right plot: 
    $N=3000$, $Z=1/15$. For small $Z$ and large $N$ the revival occurs after a time that depends only on the single parameter $NZ$.}
\end{figure}

\section{Final remarks}

It would be interesting to analyze the other experiments involving finite-level atoms, especially those with masers and mazers \cite{Walther1,Walther2,Walther3,Lamb} but a technical difficulty is that exact solutions are not there available at the moment. However, the experiments testing spectra of light 
\cite{Kimble1,Kimble3,Kimble4,Kimble5,Maunz} 
in cavity QED are another realistic goal in this context. We have already computed the vacuum Rabi splitting, with the conclusion that for $N\to\infty$ we reconstruct the standard results, which is another example of the correspondence principle. The work on comparison of the theory with experiment is in progress, and we will present the results in a separate paper. 

The structure of vacuum collapses and revivals is like a fingerprint of the representation. The parameter $NZ$ determines the distance in time between the reviving peaks. For physical reasons $Z$ must be a very small nonzero number, and thus $N$ has to be very large, although finite. Confirmation that $N<\infty$ is physical would have fundamental consequences for renormalization theory, vacuum energy with all its implications, and studies of entanglement in cavity QED.  The correspondence principle  turns $N<\infty$ theories into generalizations of standard quantum optics. Our discussion explains why it is very unlikely that $N<\infty$ can be found inconsistent with experiment. And this is interesting in itself.

We are indebted to M.~Brune for the data, and Y.-T. Chough, D.~R.~Finkelstein, S.~Haroche, J.~Naudts, G.~Nogues, and W.~Schleich for various comments.
This work was done as a part of the Polish Ministry of Scientific Research and Information Technology (solicited) project PZB-MIN 008/P03/2003.


\begin{thebibliography}{}
\bibitem{Q}Different approaches to quantization involve local \cite{Q0} and nonlocal \cite{Efimov} fields, or path-intergral \cite{Q1}, geometric \cite{Q2}, twistor \cite{Q3}, star-product \cite{Q4}, quantum-logic \cite{Fink}, and coherent-state quantizations \cite{Q5,Q51}, to name a few. The approach we advocate, based on reducible representations, is closer to what Finkelstein calls general quantization, or regularization by quantization \cite{Q6}. In fact, this is precisely an example of quantization by means of generalized covariance systems \cite{Q7}.
\bibitem{nCCR}$N=1$ representations occur if in the harmonic oscillator energy formula 
$E_n=\hbar\omega(n+1/2)$ the number $\omega$ is not a parameter but an eigenvalue of some operator, that is, a quantum number. The representations with $N>1$ correspond to a gas consisting of $N$ bosonic noninteracting oscillators of such a type, and vacuum is their zero-temperature Bose-Einstein condensate. Predictions of the standard theory are reconstructed in the weak limit 
$N\to\infty$, but finite-time experiments cannot distinguish between $N=\infty$ and $N$ finite but large.
\bibitem{cut-off}Let us note that the cut-off occurs neither in the Hamiltonian nor in the CCR  commutators. In effect, when we compute eigenvalues of the Hamiltonian we will not find there any dependence on $Z_k$. This observation is important since it shows that our 
`regularization by quantization' (in the sense of \cite{Q6}) is mathematically completely different from the usual `regularization by cut-off', in spite of the fact that for $N\to\infty$ they look very similar at the level of averages. This similarity between the two ways of regularizing may be very confusing and often leads to misunderstandings. Perhaps, the simplest exercise that may help to see the difference is to compute the spectrum of the free Hamiltonian, compare it with average energy, and understand why the vacuum energy is finite for finite $N$, even though the set of frequencies is infinite.
\end{thebibliography}
\end{document}